\documentclass[journal]{IEEEtran}
\usepackage[pdftex]{graphicx}
\usepackage{amsmath}
\usepackage{amsfonts}

\usepackage{threeparttable}
\usepackage{booktabs}

\usepackage{algorithmic}
\usepackage[linesnumbered, ruled, vlined]{algorithm2e}
\SetAlFnt{\small}
\SetEndCharOfAlgoLine{}
\usepackage{xcolor}

\SetCommentSty{myComment}
\usepackage{cite}
\usepackage[hyphens]{url}

\usepackage{hyperref}

\usepackage{soul}
\soulregister\cite7
\soulregister\citep7
\soulregister\citet7
\soulregister\ref7
\soulregister\pageref7

\begin{document}
\title{Review of Learning-Assisted Power System Optimization
	\thanks{This work was supported in part by the National Key R\&D Program of China (No. 2020YFB0905900) and National Natural Science Foundation of China (No. 51777102, No. U1766212).}
	\thanks{G. Ruan, H. Zhong, G. Zhang, Y. He and X. Wang are with the State Key Lab of Power Systems, Department of Electrical Engineering, Tsinghua University, Beijing 100084, China. T. Pu is with the China Electric Power Research Institute, Beijing  100192, China.}
	\thanks{Corresponding author: H. Zhong (zhonghw@mail.tsinghua.edu.cn).}
}%
	
\author{
	Guangchun~Ruan,~\IEEEmembership{Student~Member,~IEEE,}
	Haiwang~Zhong,~\IEEEmembership{Senior~Member,~IEEE,} 
	Guanglun~Zhang,~\IEEEmembership{Student~Member,~IEEE,}
	Yiliu~He,~\IEEEmembership{Student~Member,~IEEE,}
	Xuan~Wang,~\IEEEmembership{Member,~IEEE,}
	Tianjiao~Pu,~\IEEEmembership{Senior~Member,~IEEE} \\
}
	
\markboth{}%
{Shell \MakeLowercase{\textit{et al.}}: Journal}
\maketitle
	
\begin{abstract} 
    With dramatic breakthroughs in recent years, machine learning is showing great potential to upgrade the toolbox for power system optimization. Understanding the strength and limitation of machine learning approaches is crucial to decide when and how to deploy them to boost the optimization performance. This paper pays special attention to the coordination between machine learning approaches and optimization models, and carefully evaluates how such data-driven analysis may improve the rule-based optimization. 
	The typical references are selected and categorized into four groups: the boundary parameter improvement, the optimization option selection, the surrogate model, and the hybrid model. This taxonomy provides a novel perspective to elaborate the latest research progress and development. We further compare the design patterns of different categories, and discuss several key challenges and opportunities as well. Deep integration between machine learning approaches and optimization models is expected to become the most promising technical trend.
\end{abstract} 
	
\begin{IEEEkeywords} 
	Smart grid, machine learning, deep learning, neural network, data-driven, artificial intelligence
\end{IEEEkeywords}

\section{Introduction} \label{SEC-INTRO} 
\subsection{Background}
\IEEEPARstart{W}{ith} the advanced computing systems and big data, machine learning has successfully entered a booming period in recent years~\cite{RV5}. Machines are proven to outperform humans in more and more applications, but most experts claim that machine learning is still fast developing and has not yet reached its peak.

Optimization is popular in tremendous applications of power system, and the typical optimization tasks include: optimal dispatch, planning, system identification, dynamic security, and electricity market operations. For these applications, the major challenge is how to design an efficient and reliable method to meet the increasing requirements for optimization performance. Conventional optimization methods have shown their limitations in  complicated and volatile environments, such as future power grids with a high penetration of renewable energy~\cite{RN13}. These conventional methods tend to repeatedly solve similar problems without accumulating any experience. Machine learning, in contrast, is powerful to gain experience from historical data and previous decisions~\cite{RN26}. Empirical studies have shown that the integration of machine learning and power system optimization is able to offer significant benefits~\cite{RN19}.
	
\subsection{Bibliometric Analysis}
We conduct a bibliometric analysis on the publications that are indexed in the well-known ``Web of Science'' database. This is helpful to provide an overview of the research trend. Here, the searching query is formulated as follows: 
	
\vspace{5pt}
\parbox{0.45\textwidth}{
\texttt{\noindent
TS=(("power system" OR "smart grid") AND ("optimization" OR "optimal") AND ("data-driven" OR "artificial intelligence" OR "machine learning" OR "deep learning" OR "reinforcement learning" OR "neural network" OR "support vector machine" OR "decision tree")). 
\vspace{6pt}
}}

\begin{figure}[b]
	\centering
	\includegraphics[width=0.98\linewidth]{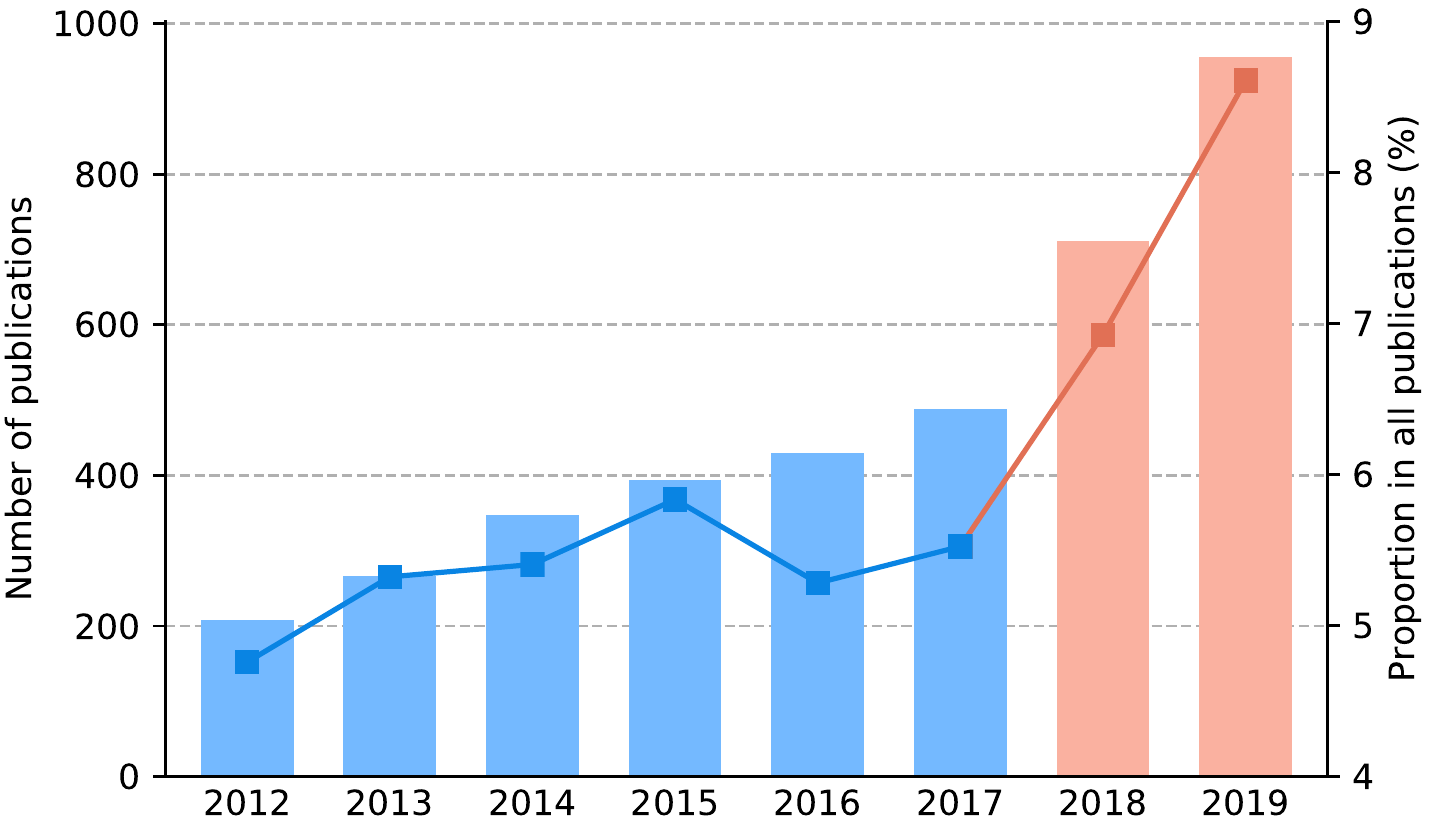}
	\caption{Research trend of learning-assisted power system optimization in recent years. The number of publications is shown with a bar chart, and the associated proportion of all publications is plotted by a line chart. A significant increase can be observed in 2018 and 2019, indicating a promising research direction. Data Source: Web of Science.}
	\label{fig_overview}
\end{figure}

Within the topics of interest, Fig.~\ref{fig_overview} shows the number of publications as well as the proportion among all publications. To derive this proportion, the number of publications on learning-assisted power system optimization is divided by the number of all power system optimization publications. To count the total number, the query expressions behind the second keyword "AND" are accordingly dropped.
	
As shown in Fig.~\ref{fig_overview}, the publications of interest only account for a small proportion from 2012 to 2017, but in the next two years, the proportion grows rapidly to 6.93\% and 8.61\%, respectively. This observation indicates that, in recent years, increasing attention is given to machine learning applications in power system optimization, and the great potential of learning-assisted power system optimization is still under exploration. 

Early researches made some preliminary attempts at Hopfield network~\cite{RN118}, radial basis network~\cite{RN115}, and self-organizing network~\cite{RN116}, but after 2010, deep learning~\cite{RN117} attracted considerable attentions and gradually became the mainstream method.

This paper pays special attention to the latest developments, and we carefully select a few articles that are the most representative work in the latest three years---24 articles are published in the first half of 2020, and 38 articles are published in either 2018 or 2019. Moreover, these articles are mainly chosen from several prestigious journals in power system domain, including \textit{IEEE Transactions on Power System, IEEE Transactions on Smart Grid}, and \textit{Applied Energy}.
	
\subsection{Comparison with Related Review Articles}
This paper shows two essential differences from the existing review articles. First, a different taxonomy is developed in this paper to demonstrate and highlight the methodological features of existing literature. In comparison, existing reviews are more focused on different applications~\cite{RV3,RV6,RV10,RV11,RV12,RV13} or different learning methods~\cite{RV1}.

Second, this paper limits the scope of references within power system optimization rather than a wide range of machine learning applications. More technical details and discussions are therefore provided in this paper to understand the most typical methodologies.

In summary, this paper intends to provide a novel taxonomy to understand how machine learning approaches may benefit the power system optimization. The selected references will be divided into different categories according to different methodological features. We further discuss the key challenges in practical applications, and recommend some potential solutions as well.

\subsection{Contributions and Paper Structure}
The major contributions are summarized as follows.
\begin{enumerate}
	\item We propose a novel and well-designed taxonomy according to different methodological features. This taxonomy helps elaborate the coordination between machine learning approaches and power system optimization models. We also summarize the key technologies for each category from a number of recent publications.
	
	\item Major challenges and opportunities in learning-assisted power system optimization are fully discussed, and the latest academic explorations are summarized to provide some possible solutions.
\end{enumerate}
	
In the rest of this paper, Section~\ref{SEC-TAX} proposes a well-designed taxonomy, and the key technologies for each category are studied in Section~\ref{SEC-TECH}. We further discuss the major challenges and future opportunities in Section~\ref{SEC-CHL}. At last, Section~\ref{SEC-CONCL} concludes this paper.

\begin{figure*}
    \centering
    \includegraphics{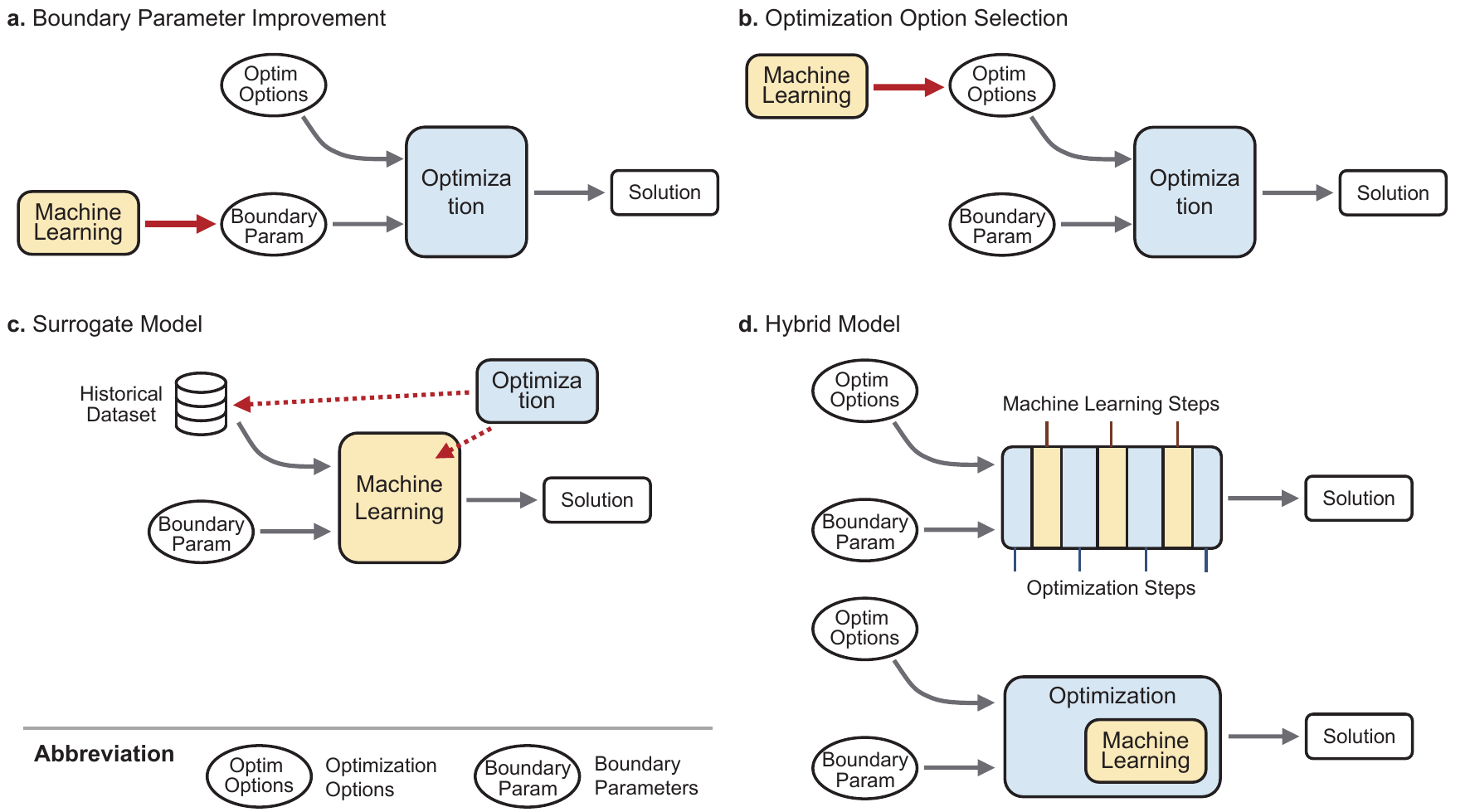}
    \caption{Overview of the proposed taxonomy. This taxonomy is focused on the methodological features of existing literature, and the main ideas of each category is graphically illustrated. This novel taxonomy contributes to understanding the coordination between machine learning approaches and optimization models.}
    \label{fig-framework}
\end{figure*}

\section{Taxonomy} \label{SEC-TAX} 
This section provides a novel and well-designed taxonomy according to different methodological features. We intend to answer a key question that how the machine learning approaches may change the power system optimization. Here, the potential benefits and risks are both taken into consideration.

Specifically, we focus on the coordination between the machine learning approaches and the optimization models, and make four categories to extract valuable insights from the existing literature. Fig.~\ref{fig-framework} shows an overview of these four categories: a)~\textbf{boundary parameter improvement}, b)~\textbf{optimization option selection}, c)~\textbf{surrogate model}, and d)~\textbf{hybrid model}.

Let us revisit the power system optimization models from a full-cycle perspective. Generally, an optimization model contains an objective function and several constraints, whose coefficients (boundary parameters) are important for a high-quality solution. Classical optimization methods are widely used, and some optimization options, e.g., initial value or iterative step size, need to be effectively configured in advance. Additionally, the surrogate machine learning model and well-designed hybrid model are two alternative solvers which can boost the optimization performance by the experience learned from historical data.

Based on these discussions, the basic ideas of the proposed four categories are quite straightforward. The first category, boundary parameter improvement, uses machine learning approaches to improve coefficient estimation. The second category is focused on selecting better optimization options. The reinforcement learning and other special machine learning approaches are working as the surrogate models or the third category. The last category, the hybrid models, includes various hybrid analytical and data-driven frameworks. Note that the last two categories are technically different from the previous two---the coordination pattern changes from a tandem structure to some iterative, coupled or other complex structures. These new structures are more likely to perform better than using optimization or machine learning model alone.

We next analyze an example of demand response from the perspectives of different categories. Reference~\cite{RN12} (Category I) focused on the uncertainty issue of effective participation period, which was one of the critical boundary parameters. Reference~\cite{RN119} (Category II) estimated a better initial ON/OFF status of responsive appliances. While \cite{RN28} (Category III) applied reinforcement learning for optimal control, \cite{RN62} designed a hybrid framework to integrate neural network and optimization models together. Both \cite{RN28} and \cite{RN62} achieved a significant algorithmic speedup.

Beyond forecasting, machine learning approaches are showing an attractive prospect for potential applications in power system optimization. The taxonomy of this paper clearly demonstrates the current academic progress, and becomes useful to design new-style models.

\begin{figure*}
	\centering
	\includegraphics[width=0.8\linewidth]{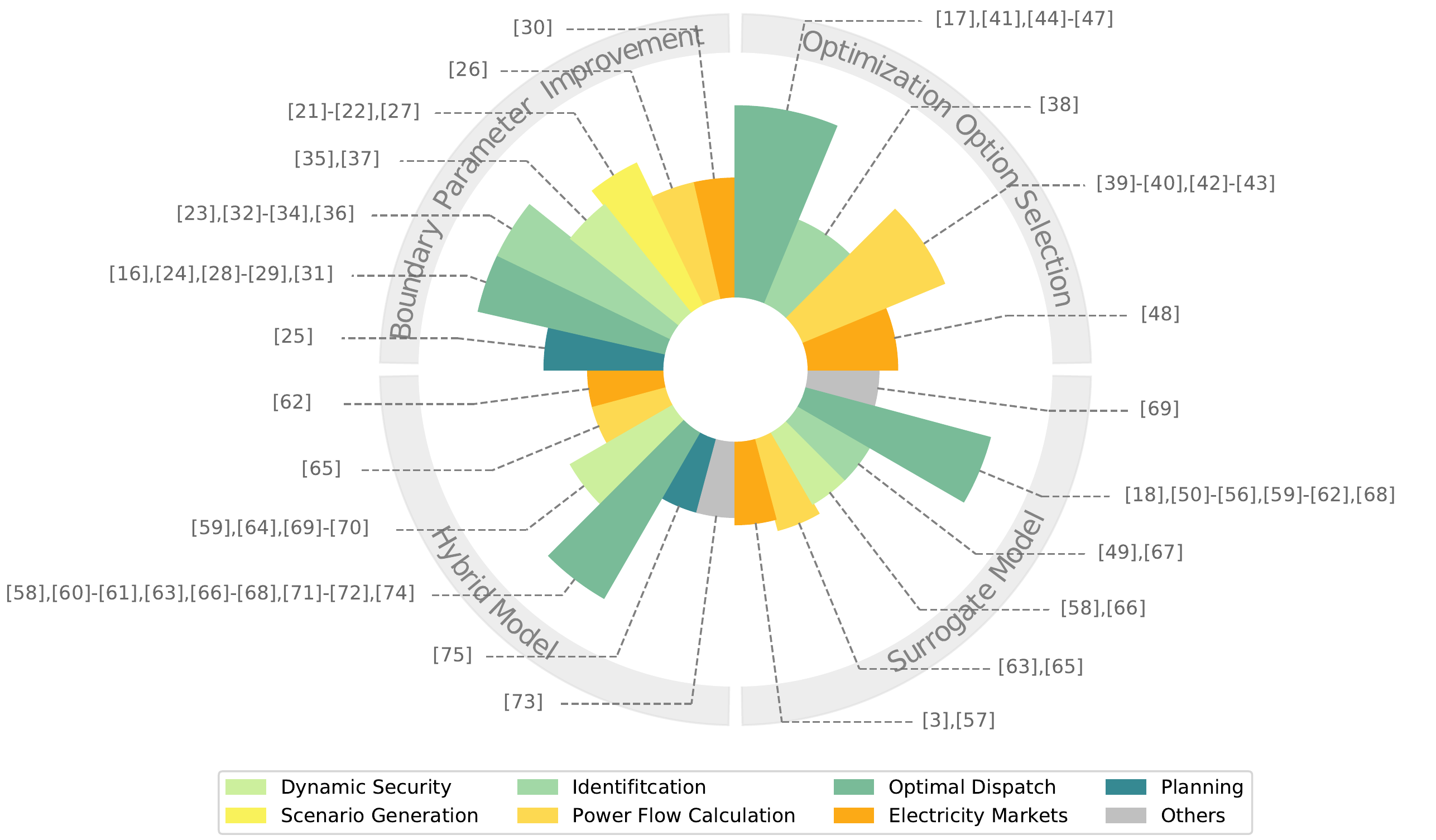}
	\caption{Overview of all selected references in Section~\ref{SEC-TECH}. These references are classified by different application scenarios and grouped by the proposed taxonomy. Different colors are used for different applications, and the height of each slice is proportional to the number of selected references. This figure is useful for a quick search by applications or the proposed taxonomy.}
	\label{fig-Ref}
\end{figure*}

\section{Key Technologies} \label{SEC-TECH} 
According to the proposed taxonomy, this section summarizes the latest research progress as well as key technologies. A list with all reference details can be downloaded from~\cite{Ref}. Fig.~\ref{fig-Ref} gives an overview of the selected references. Here, different colors are assigned according to the application scenarios, and the height of each slice is proportional to the number of selected references. We will next dive into the technical details of each category.

\subsection{Category 1. Boundary Parameter Improvement}
This category uses machine learning approaches to improve the estimations of boundary parameters, and thus formulates more accurate feasible regions and better optimal solutions. There are many factors that may deteriorate the accuracy of boundary parameters, including natural variability, human behaviors, and partial observability of power systems.

\textbf{Impact of natural variability.} With high penetration of renewable energy, power systems are meeting the challenges of increasing uncertainty. Reference~\cite{RN1} applied the generative adversarial networks to create scenarios of renewable energy, and this method was also applicable to other highly uncertain environments. Conditional generative adversarial networks performed better than classical methods to account for wind power uncertainty~\cite{RN2}. In \cite{RN7}, a deep learning approach was proposed to identify the fault types of transmission lines, and the features could be automatically learned from raw data. Reference~\cite{RN9} considered a wind and storage power plant that took part in a pool market. The authors combined the multivariate clustering technique and a recurrent neural network in order to model the uncertain electricity prices and wind power. Considering the socio-technical complexities in the planing stage, \cite{RN15} developed a data-driven framework to analyze the distributed energy resources. In~\cite{RN110}, the natural variability is captured by joint chance constraints, and the bounds are estimated using support vector machines.

\textbf{Impact of human behaviors.} Existing literature has used machine learning approaches to eliminate the randomness caused by human behaviors and obtain a better solution. Reference~\cite{RN3} proposed a long short-term memory neural network to adequately consider the uncertain prices in electricity markets. Neural networks played a significant role in balancing between the thermal comfort and energy use in buildings~\cite{RN10}, achieving optimal dispatch in ancillary services market~\cite{RN11} and reduced load curtailments~\cite{RN12}. In~\cite{RN16}, a data-driven model was used to increase the learning ability for price responsive behaviors. A game theory-based market strategy was designed to increase profit of each market participant with operational security guarantees. Reference~\cite{RN111} designed several local control strategies in distribution networks by learning historical control schemes.

\textbf{Impact of partial observability.} Partial observability may result in inaccurate models and decisions. Deep learning models were deployed to identify time-varying parameter for composite load model~\cite{RN4}, phasor measurement unit data manipulation attacks~\cite{RN5}, phase detection in power distribution systems~\cite{RN8}, and feature extraction for security rules~\cite{RN17}. To enhance the observability of distribution systems, \cite{RN6} presented a data-driven method to estimate the daily consumption patterns of the customers without smart meters. In addition, \cite{RN18} showed that generative adversarial networks could quickly assess the dynamic security with missing data.

\subsection{Category 2. Optimization Option Selection}
This category intends to boost the optimization performance by selecting better optimization options, e.g., initial values. Practice shows that some options have significant impacts on the convergence feature and speed, but the default settings are sometimes unsatisfactory. Machine learning approaches, in this context, can provide effective guidance from the past experiences.

Many researches use machine learning approaches to estimate a good initial value, which is beneficial for a warm-start algorithm. Reference~\cite{RN41} proposed a ``learn to initialize’’ strategy to improve the Gauss-Newton algorithm. The authors achieved this with a neural network, and designed a special loss function (only penalizing the maximized errors) to improve the overall performance. Reference~\cite{RN48} established a predict-and-reconstruct approach to predicting the generation states (part of the decision variables), and reconstructed the phase angles (other part of the decision variables) using power flow equations. A deep neural network was developed for this task, and the network size was properly configured according to approximation accuracy. In~\cite{RN39}, a data-driven approach to reconstructing the solution of a centralized optimal power flow was proposed. The idea was that local controllers could find a near-optimal solution by learning the limited but locally available data.

Some other extensions were discussed in~\cite{RN43, RN44, RN45}. The supervised and transfer learning were applied in~\cite{RN43} to estimate the Pareto front that is made up with a series of initial values. This task was indeed more difficult than \cite{RN41, RN48, RN39}. The numerical tests indicated that such estimation might cause large errors under specific conditions, so further validation and fine-tuning were extremely important. Reference~\cite{RN44} proposed a linear power flow model to accelerate and approximate the power flow calculation. This work was further extended in~\cite{RN45} to tackle the challenge of hidden measurement noises. The authors formulated three quadratic programming models with several Jacobian-matrix-guided constraints to achieve this goal.

The potential advantages in discrete optimization are more attractive than expected. An early work~\cite{RN49} introduced a combined approach for  unit commitment problems. This approach first used a neural network to determine the discrete variables, and after that, applied the simulated annealing method to optimize the continuous variables. Case studies showed that the neural network found near-optimal commitment results, and achieved a roughly 50x computing speedup. A recent work~\cite{RN47} made further progress in security-constrained unit commitment. Different machine learning approaches have been adopted to study previously solved instances, and accelerate the computation by predicting redundant constraints, good initial feasible solutions, and active affine subspaces. On average, the authors achieved a 4.3x speedup with optimality guarantees and a 10.2x speedup without optimality guarantees (but with no discernible solution difference). It provided a valuable insight that predicting warm start states or active hyper-planes were significantly harder than estimating redundant constraints. Similar technique was also applicable in demand response. Reference~\cite{RN119} used a neural network to estimate the optimal ON/OFF status of home appliances, which could be regarded as an efficient warm start setting.

Machine learning approaches were also useful to configure other optimization options. Reference~\cite{RN46} created an effective algorithm selector by machine learning approaches and found less overloads in power flow management. Similar method was also adopted in a unit commitment problem~\cite{RN42} where a learning model was trained to assign weights according to some heuristic rules. Reference~\cite{RN40} formulated a three-stage framework (mid-term, short-term, and real-time) for outage scheduling, and a nearest neighboring classifier was trained to approximate the intermediate results to speed up the mid-term decision-making.

\begin{figure}[t]
	\centering
	\includegraphics[width=0.95\linewidth]{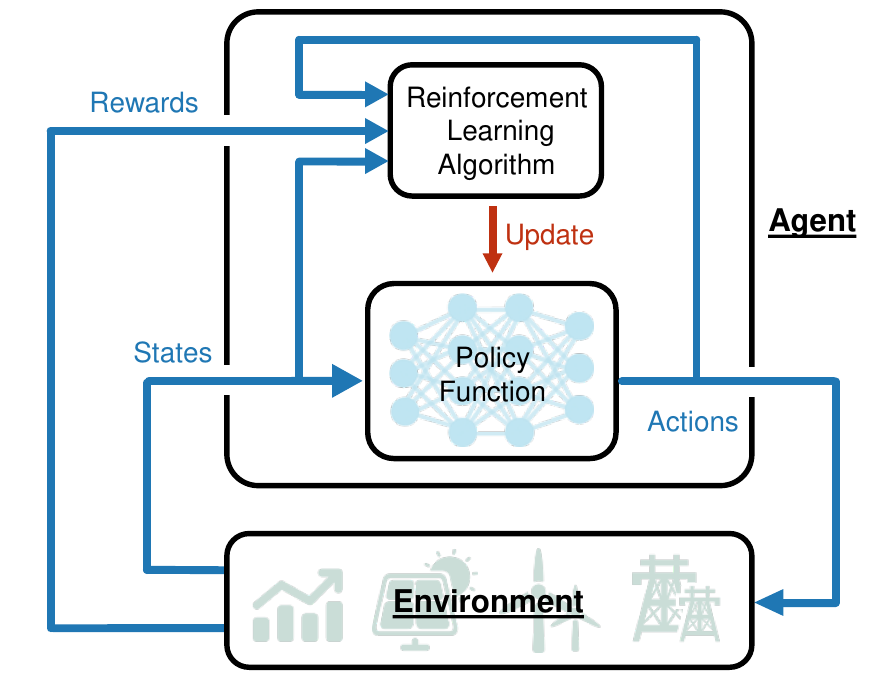}
	\caption{Basic structure of reinforcement learning. The agent can observe the states of the environment, and take actions in response. After that, the environment gives back a reward according to the actions. The major objective of the agent is maximizing its total reward during a decision period.}
	\label{fig-RL}
\end{figure}

\subsection{Category 3. Surrogate Model}
This category seeks to completely replace optimization models by other data-driven models. These surrogate models are extremely powerful when analytical models are unavailable or too computationally expensive. We study the most prevalent method, reinforcement learning, as well as other specific methods in the following paragraphs.

Reinforcement learning is developed for sequential decision making that can be formulated as Markov decision process. It becomes more and more popular as a surrogate method for various complex optimization or control tasks. The basic idea is to construct an agent that takes actions to maximize the cumulative rewards, and the objective is to design an optimal policy for action selection. Many reinforcement learning algorithms have been developed with diverse reward estimations and policy designs. Additionally, deep reinforcement learning further enhanced the intelligent ability of agents to adapt to different applications. Fig.~\ref{fig-RL} shows the basic structure of (deep) reinforcement learning.

A major advantage of reinforcement learning is the model-free formulation that is independent of prior knowledge.Reference ~\cite{RN26} used deep reinforcement learning to study the responsive behaviors of the microgrids in a distribution network. Reference~\cite{RN25} applied fitted Q-iteration to control a cluster of electro-thermal loads with unknown characteristics. A convolutional neural network was constructed to process the high-dimensional inputs and capture the hidden patterns. Reference~\cite{Rh6} applied reinforcement learning to control a group of heterogeneous batteries, and their diverse physical characteristics were exploited to improve the overall efficiency. A coordinated wide-area voltage controller was designed in \cite{Rh7} to optimize the transient process of an electric grid, showing a better dynamic stability than conventional methods. Reference~\cite{Rh8} formulated an empirical model to approximate the degradation probabilities of grid components, and applied reinforcement learning to make maintenance decisions. Reference~\cite{RN27} used deep reinforcement learning to navigate the electric vehicles that need recharging, and the total travel time and cost were greatly minimized.

Another advantage of reinforcement learning is its online learning ability to make immediate response to the fluctuations of exogenous factors. For example, \cite{Rh1} solved online optimization to minimize energy cost and users’ dissatisfaction with regard to the electricity prices and solar output. Reference~\cite{RN28} used an artificial neural network to predict day-ahead prices, and combined it with a Q-learning model for optimal dispatch of home energy system. Similar methods were also applied to microgrid scheduling, e.g., \cite{Rh2,Rh3}.

Reinforcement learning is also useful in distributed operation or under information asymmetry. Reference~\cite{RN34} designed a multi-agent system to represent independent generation companies. A policy gradient method was applied to make decisions with limited market information. For current and voltage control, \cite{RN22} integrate the consensus method and deep reinforcement learning to coordinate distributed generators in an island microgrid. The distributed reactive power optimization was solved by the collaborative equilibrium Q-learning to minimize operating cost and carbon emission~\cite{RN21}. Reference~\cite{RN20} proposed a multi-agent architecture to schedule electric vehicle charging based on Q-learning and W-learning.

A major concern of reinforcement learning is that the black box formulation can hardly consider physical characteristics or constraints. Large decision errors may happen in some unexpected cases because the performance of reinforcement learning heavily depends on the training data.

To solve this problem, safe reinforcement learning was developed to guarantee certain security constraints. The basic idea is to introduce some penalty terms corresponding to the security constraints, and minimize them in priority during the learning process. Reference~\cite{RN23} adopted this idea to consider charging constraints of electric vehicle batteries.Reference~\cite{RN24} optimized voltage and reactive power by a safe off-policy deep reinforcement learning algorithm to avoid voltage violations.

Some studies have also explored embedding physical characteristics within deep learning methods. A physics-guided neural network approach was proposed in~\cite{RN36} to calculate probabilistic power flow. The training process was improved by the grid characteristics, and the case study showed a great computation speedup. Similar applications were completed by \cite{RN112,RN113} with a fully connected neural network and extreme learning machine, respectively. A graph convolutional network was trained in \cite{RN37} to capture the topology information, and calculate the optimal load-shedding under contingency. Reference~\cite{RN38} used a convex neural network to approximate the electrothermal characteristics of a building, and then solve a convex optimization to minimize the electricity consumption. In the online control of plug-in electric bus~\cite{Rh5}, the trip information was represented by the length ratio, and this mapping relationship was learned by a neural network. Reference~\cite{RN35} studied the pump speed control to optimize the waste heat recovery of internal-combustion engines.The authors used the dynamic programming and supervised learning methods, whose inputs were specially designed according to several physical differential equations.

\begin{figure}[t]
	\centering
	\includegraphics[width=1\linewidth]{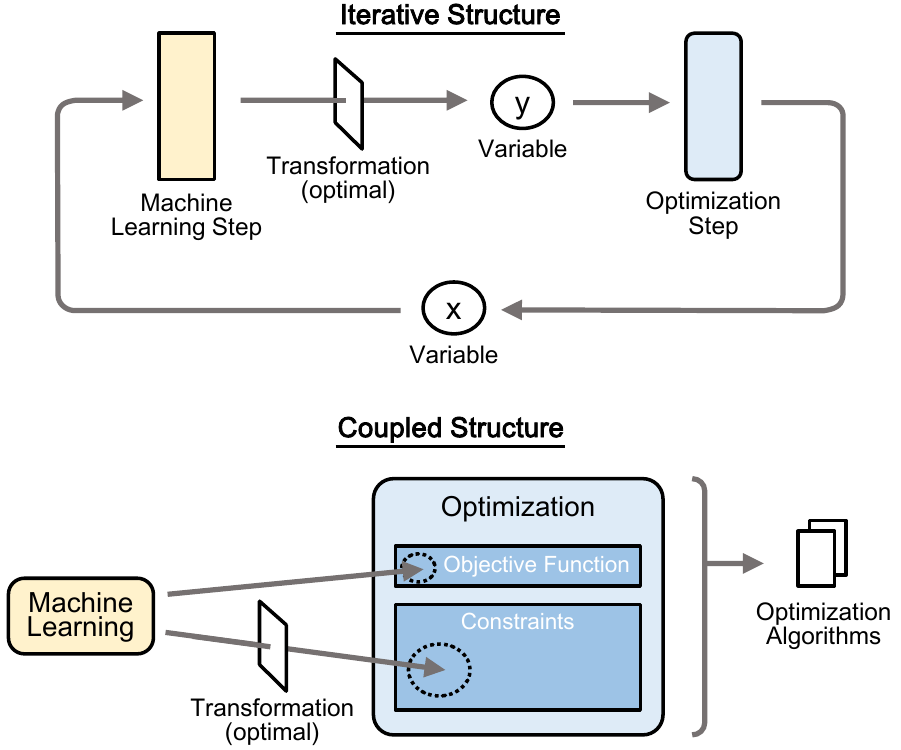}
	\caption{Structure details of two typical hybrid models. An iterative structure takes learning steps and optimization steps alternately, while a coupled structure replaces certain inaccurate parts in an optimization model with machine learning formulations.}
	\label{fig-cat4}
\end{figure}

\subsection{Category 4. Hybrid Model}
This category combines the machine learning approaches and optimization models together to boost the overall performance. There are two hybrid types shown in Fig.~\ref{fig-framework}(d), and more details are provided in Fig.~\ref{fig-cat4} as well. Here, one type is an iterative structure with machine learning and optimization steps, and the other is a coupled structure that embeds machine learning models into optimization models to replace some inefficient parts. Such combinations are expected to provide a deep integration between two kinds of models, and fully explore their hidden potentials. Some case studies have achieved higher modeling accuracy and optimality property at the same time, e.g.,~\cite{RN62, RN64}.

\textbf{Iterative structure.} This structure is very popular in many optimization and optimal control tasks, and the typical flowchart is shown in Fig.~\ref{fig-cat4}. Reference~\cite{RN62} proposed an accelerated algorithm for distributed demand response, which applied a neural network to iteratively predict the consumers’ price responses. The authors further developed a transformation model to search for better step sizes. The most promising feature of this algorithm was that it could cut off 60-80\% iterations but still guarantee optimality property. In addition to step sizes, \cite{RN63} showed that neural network was also able to improve searching directions of a sequential linear programming model. This paper considered an information asymmetric situation where a retailer company needed to make decisions with limited knowledge of its consumers. The dual neural networks were designed and one network was working to transform and derive searching directions. 

Other intelligent optimization algorithms were also implemented in existing researches. Reference~\cite{RN51, RN50} formulated a similar optimal dynamic pricing model for retailers but solved the model by the genetic algorithm and mean-variance mapping optimization, respectively. Here, the consumer responsive features were learned and coordinated with the intelligent optimization algorithms later. Reference~\cite{RN59} established a three-stage model predictive control where neural networks were predicting the energy demand and renewable energy supply to construct the optimal decisions. In~\cite{RN65}, a hybrid model and data-driven simulation platform was operating in real-time for selecting the power system security features. Within this platform, the analytical models generated the samples, and they were then analyzed to extract several fine-tuned security rules.

\textbf{Coupled structure.} This structure is more complex than the iterative structure mentioned above. As shown in Fig.~\ref{fig-cat4}, current researches made effort on embedding machine learning models in the objective function or constraints, and the key technical difficulty is how to design the optimization algorithms and transformation process. 
Reference~\cite{RN68} extracted a mapping function from the neural network formulation and included it as a constraint in an optimal power flow model. This hybrid model was then solved by a nonlinear programming optimizer. Reference~\cite{RN52, RN60} integrated the machine learning models in the constraints of a building energy optimization model. To solve this model, \cite{RN52} applied particle swarm optimization, while \cite{RN53} applied a hybrid approach of exhaustive search method and subsequent quadratic programming. Similarly, the particle swarm optimization was also conducted in~\cite{RN54} to solve an optimization whose constraints were modeled by a radial basis function neural network. Reference~\cite{RN56} formulated a preventive control model with a Bayesian neural network that could predict the steady-states. This model was later solved by Bayesian optimization.

We next talk about some typical transformation measures. A piece-wise linear approximation transformed neural networks in~\cite{RN57} to finally derive a mixed integer programming model. Reference~\cite{RN64, RN61} chose different machine learning approaches as an alternative. In~\cite{RN64}, the sparse oblique decision tree was applied to learn some accurate, understandable, and linear security rules for economic dispatch. These rules could be embedded in an optimization as several mixed integer linear constraints. In~\cite{RN61}, the authors conducted an extreme learning machine to enhance the hydrostatic tidal turbine control. This model was basically a linear model and can be easily transformed to linear constraints. Another special technique was introduced in~\cite{RN67}, where the authors designed a sequential approximation method with dynamically trained neural networks. With the consideration of running time, such dynamic training might be more suitable for small networks.

Many researches have formulated objective functions with machine learning elements. Reference~\cite{RN58} trained a neural network to learn the combined heat and power simulation results, and formulated the objective function with this neural network integrated. The dispatch schedule was optimized by genetic algorithm. Reference~\cite{RN55} designed a convolutional-neural-network-based classifier for faulted line localization. The placement problem of phasor measurement units was a hybrid model, and the objective function contained the loss function of convolutional neural network. This problem was actually a special hyperparameter optimization, and was later solved by greedy algorithm.

\subsection{Comparison and Comments}
Overall, the selected references have very diverse ideas and features. We will compare all above categories and make further comments on their applications.

\textbf{Difficulty of the learning tasks.} Considering the same data dimension and  precision requirement, the Category~3 contains the most difficult learning tasks. Here, the so-called ``difficulty’’ can be measured by how large a machine learning model is needed to finish the task. The Category~2 usually takes the second place, but the situations for the remaining two categories are uncertain.

Regardless of some exceptions, the surrogate models (Category~3), especially those reinforcement learning models, are usually very complicated to calibrate, so a large amount of data and computing resources are needed. These models might meet great difficulty if the simulation data are not available or effective. Furthermore, the boundary parameter improvement (Category~1) is often easier to achieve, but its potential benefits are strongly related to the specific cases. As for the remaining two categories, some physical knowledge may be relatively helpful to boost the overall performance.

\textbf{Difficulty of the model designs.} For a similar task, the Category~4 or Category~2 methods are often in need of more efforts or even case-by-case designs. In contrast, the Category~3 is more likely to have some off-the-shelf tools, and little manual processing is needed.

There are two significant features in power system optimization---highly sensitive to decision errors, and dependent on extensive physical knowledge. However, leveraging these physical knowledge to design a dedicated framework or machine learning model is still under exploration. In this aspect, the hybrid model (Category~4) and other variants in other categories deserve more exploration although more design efforts are needed.

\section{Challenges and Opportunities} \label{SEC-CHL} 
Power system optimization always has high requirements for the algorithm reliability because a potential failure could lead to a significant financial loss. However, machine learning is still having troubles in real-world power system applications, and the major concerns should be carefully analyzed and handled~\cite{RN120}. With this purpose, this section points out three major challenges, and shares some possible solutions as well.

\subsection{Data Bottleneck}
Collecting clean and reliable data is essential to every machine learning application, and the requirement for modern deep learning models is even higher. Two special features of power system data should be noted here: First, due to privacy concerns or confidentiality requirements~\cite{RV1}, very few real-world data sets are public available, simulation data are therefore widely applied as an alternative~\cite{RV2}. Second, real-world data are often imbalanced, and the rare part may be extremely important, e.g., unstable system conditions~\cite{RV10}.

Data issues show adverse impacts on all four categories, and thus become the major bottleneck for real-world applications. Apparently, it will bring more risks to those data-intensive models In addition to expanding data sets with policy support, there are also some emerging technologies that could help tackle these data issues---data augmentation and few-shot learning.

Data augmentation is a strategy that significantly increases data volume by a series of transform operations. Reference~\cite{RN102} collected a list of useful resources, including classical techniques, papers and Github repositories. For example, the mainstream techniques for time-series data augmentation can be classified into simple operations (warping, jittering and perturbing) and advanced operations (embedding space and generative approaches).

Few-shot learning intends to feed machine learning models with limited amount of data. The basic idea is to use prior knowledge to avoid the unreliable performance of empirical risk minimizer. In this aspect, the authors of~\cite{RN103} reviewed some model-based methods (constrain the model complexity) and algorithm-based methods (constrain the search strategy for optimal parameters).

\subsection{Robustness and Prediction Errors}
Power system optimization imposes high requirements on accuracy as well as robustness. In this context, it is crucial to take care of the robust vulnerability---In extreme cases, a small input change may lead to a significant drop in accuracy.

There are two perspectives to analyze and understand the robustness issue:
\begin{itemize}
	\item What is the output change of machine learning models with a fluctuation in model inputs?
	\item What is the change of optimal solution with a fluctuation in machine learning model outputs?
\end{itemize}
For Category~3, these two perspectives merge into one, while for other categories, the second perspective is important but seldom discussed. We will next introduce some latest research explorations in the two perspectives.

From the first perspective, adversarial examples are fairly helpful to examine the robustness of machine learning approaches. Reference~\cite{RN104} studied the worst-case adversarial perturbations and found the robustness might be badly harmed. A recent work~\cite{RN101} argued that, in fact, these adversarial examples were features rather than bugs. The authors further demonstrated the robust and non-robust features, and the non-robust ones were the main cause for robust vulnerability.

\begin{figure}[t]
	\centering
	\includegraphics[width=0.94\linewidth]{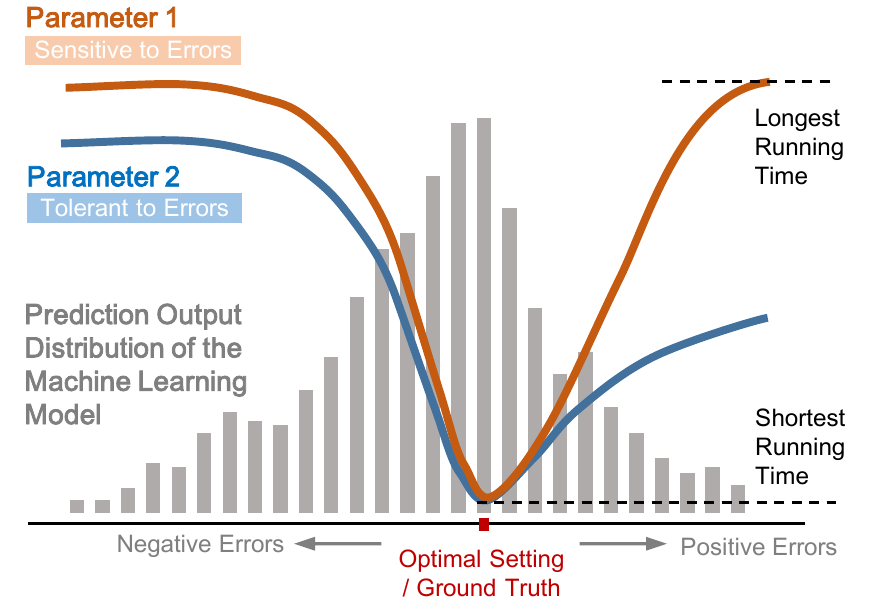}
	\caption{Comparison between error-tolerant parameters and error-sensitive parameters. In this example, error-tolerant parameters can robustly ensure a shorter running time than error-sensitive parameters, even though their best performances are nearly the same.}
	\label{fig-pred-error}
\end{figure}

The second perspective is highly related to the specific characteristics of optimization model. We can divide the model parameters and optimization options into two parts: error-tolerant parameters and error-sensitive parameters. Fig.~\ref{fig-pred-error} gives an illustrative example to show the difference. Our main focus is on the overall optimization performance when machine learning models make positive and negative prediction errors. It is shown that an error-tolerant parameter can robustly guarantee a shorter running time for optimization. For Category~1, 2 and 4, choosing error-tolerant parameters to coordinate between machine learning approaches and optimization models could improve the robustness of the whole system.

\subsection{Interpretability}
Interpretability describes how much people can understand the decisions made by machine learning approaches. Many machine learning models, e.g., neural networks, are widely regarded as ``black box’’ models~\cite{RV1}. Ensemble learning makes the interpretability even worse by combining several black box models together. As a consequence, low interpretability severely hinders the wide applications of machine learning approaches in the power industry. 

There are two perspectives of the interpretability:
\begin{itemize}
	\item How are the optimal weights found by machine learning models?
	\item How can machine learning model outputs improve the optimization performance?
\end{itemize}
Similar as the previous subsection, for Category~3, these two perspectives merge into one, while for Category~1, the second perspective is very intuitive---closer to the ground truth is better. In the remaining two categories, both two perspectives are important and should be carefully discussed.

\begin{figure}[t]
	\centering
	\includegraphics[width=1\linewidth]{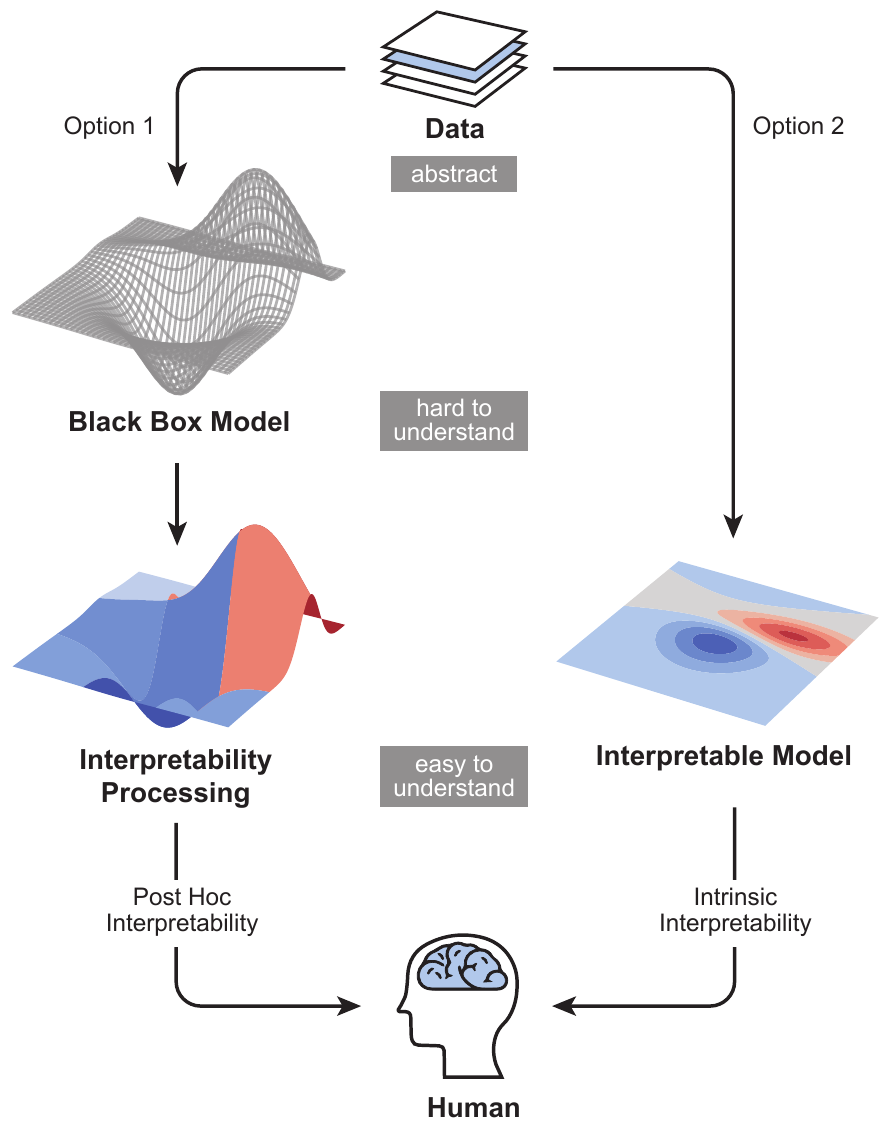}
	\caption{Two options to achieve machine learning interpretability. These two options, applying an interpretable model or making further processing on a black box model, are illustrated to translate the original data to some easy-to-understand explanations.}
	\label{fig-interp}
\end{figure}

The first perspective is a conventional topic that have been discussed for years in machine learning community. Reference~\cite{RN100} introduced most of the important progresses in this domain which are also shown in Fig.~\ref{fig-interp}. Basically, there are two options to achieve interpretability: applying an interpretable model or making further processing on a black box model. Model-agnostic interpretation methods, with rapid advances in recent years, are able to extract many human-friendly features and visualization results.

The second perspective is dedicated for learning-assisted optimization, and can be probably demonstrated with the help of optimization theory. There often exist a set of optimal configurations to boost the optimization performance, but they may be computing-intensive or even intractable. Machine learning approaches, however, can approximate these configurations efficiently. A typical example is the selection of near-optimal step sizes in~\cite{RN62}.

\section{Conclusion} \label{SEC-CONCL} 
This paper conducts a comprehensive review of learning-assisted power system optimization. A novel and well-designed taxonomy is proposed in this paper to categorize the existing articles by their methodological features. The latest research progress and key technologies are thoroughly summarized, along with further comments on the key challenges in real-world applications.

We strongly realize that the deep integration of machine learning and power system optimization is a promising future trend. This review is expected to offer some useful information and new insights in this domain.

\bibliographystyle{ieeetr}
\bibliography{refs}

\end{document}